\documentclass{pnastwo}

\usepackage{graphicx}
\usepackage{amsmath}
\usepackage{amsfonts}
\usepackage{amssymb}
\usepackage{xcolor}
\usepackage{bm}
\usepackage{psfrag}
\usepackage{subfigure}
\usepackage{cite}
\usepackage{setspace}

\begin{document}
\newcommand{\bee}{\begin{equation}}
\newcommand{\eee}{\end{equation}}
\newcommand{\ba}{\begin{eqnarray}}
\newcommand{\ea}{\end{eqnarray}}

\title{Why glass elasticity affects the thermodynamics and  fragility of super-cooled liquids}
\author{Le Yan, 
 Gustavo D\"uring, 
 \and Matthieu Wyart\affil{1}{Center for Soft Matter Research, Department of Physics, New York University, New York, NY 10002, USA}}

\maketitle


\begin{article}

\begin{abstract}

Super-cooled liquids are characterized by their fragility:  the slowing down of the dynamics under cooling is more sudden and the jump of specific heat at the glass transition is generally larger in fragile liquids than in strong ones. Despite the importance of this quantity in classifying liquids, explaining what aspects of the microscopic structure controls fragility remains a challenge. 
Surprisingly, experiments indicate that the linear elasticity of the glass -- a purely local property of the free energy landscape -- is a good predictor of fragility.  In particular, materials presenting a large excess of  soft elastic modes, the so-called boson peak, are strong. This is also the case for  network liquids near the rigidity percolation, known to affect elasticity. Here we introduce a model of the glass transition based on the assumption that particles can organize locally into distinct configurations,  which are coupled  spatially via elasticity. The model captures the mentioned observations connecting elasticity and fragility. We find that materials presenting an abundance of soft elastic modes have little elastic frustration: energy is insensitive to most  directions in phase space, leading to a small jump of specific heat. In this framework  strong liquids turn out to lie the closest to a critical point associated with a rigidity or jamming transition, and their thermodynamic properties are related to the problem of number partitioning and to Hopfield nets in the limit of small memory.

\end{abstract}




\section{Introduction}

When a liquid is cooled rapidly to avoid crystallization, its viscosity increases up to the glass transition where the material becomes solid. Although this phenomenon was already  used in ancient times to mold artifacts, the nature of the glass transition and the microscopic cause for the slowing down of the dynamics 
remain controversial. Glass-forming liquids are characterized by their fragility \cite{Ediger,stillinger}: the least fragile liquids are called strong, and their characteristic time scale $\tau$
follows approximatively an Arrhenius law $\tau(T)\sim \exp(E_a/k_BT)$, where the activation energy $E_a$ is independent of temperature. Instead in fragile liquids the activation energy 
grows as the temperature decreases, leading to a sudden slowing-down of the dynamics. The fragility of liquids strongly correlates with their thermodynamic properties \cite{wang,martinez}: the jump in the specific heat that characterizes the glass transition is large in fragile liquids and moderate in strong ones. Various theoretical works \cite{adam,woly,BB,dyre}, starting with Adam and Gibbs, have proposed explanations for such correlations. By contrast few propositions, see e.g. \cite{Wolynes,lubchenko,tanaka}, have been made to understand which aspects of the microscopic structure of a liquid determines its fragility and the amplitude of the jump in the specific heat at the transition. 

Observations indicate that the linear elasticity of the glass is a key factor determining fragility -- a fact a priori surprising since linear elasticity is a local property of the energy landscape, whereas fragility is a non-local property characterizing transition between  meta-stables states. In particular (i) glasses are known to present an excess of soft elastic modes with respect to Debye vibrations at low frequencies, 
the so-called boson peak  that appears in scattering measurements \cite{phillips_book}. The amplitude of the boson peak is strongly anti-correlated with fragility, both in network and molecular liquids: structures presenting an abundance of soft elastic modes tend to be strong \cite{novikov,Ngai}.
(ii) In network glasses, where particles interact via covalent bonds and via the much weaker Van der Waals interactions, the microscopic structure and the elasticity can be monitored  by changing continuously the composition of compounds \cite{ANGELLCHALCO,angell,boolchand2}. 
As the average valence $r$ is increased toward some threshold $r_c$, the covalent networks display a rigidity transition \cite{Phillips,PhillipsThorpe} where the number of covalent bonds is just sufficient to guarantee mechanical stability.  Rigidity percolation has striking effects on the thermal properties of super-cooled liquids: in its vicinity, liquids are strong \cite{angell2} and the jump of specific heat  is  small \cite{ANGELLCHALCO}; whereas they become 
fragile with a large jump in specific heat  {\it both} when the valence is increased,  and 
decreased \cite{ANGELLCHALCO,angell2}. It was argued \cite{Wolynes}  that fragility should decrease with valence, at least when the valence is small.  
There is no explanation however why increasing the valence affects the glass transition properties in a non-monotonic way, and why such properties are extremal when the covalent network acquires rigidity \cite{comment}.

Recently it has been shown that the presence of soft modes in various amorphous materials, including granular media \cite{O'Hern03,brito4,Wyart05,Wyart053}, Lennard-Jones glasses \cite{Wyart053,Xu07}, colloidal suspensions \cite{brito3,bonn1,chen} and silica glass \cite{silica1,Wyart053} was controlled by the proximity of a jamming transition\cite{revue}, a sort of rigidity transition  that occurs for example when purely repulsive particles are decompressed toward vanishing pressure \cite{O'Hern03}. Near the jamming transition spatial fluctuations play a limited role and simple theoretical arguments \cite{Wyart05,Wyart053} capture the connection between elasticity and structure. They imply that soft modes must be abundant near the transition,  suggesting a link between observations (i) and (ii). However these results apply to linear elasticity and cannot explain intrinsically non-linear phenomena such as those governing fragility or the jump of specific heat. 
In this article we propose to bridge that gap by introducing a  model for the structural relaxation in super-cooled liquids. Our starting assumption is that  particles can organize locally into distinct configurations,  which are coupled at different points 
in space via elasticity. We study what is perhaps the simplest model  realizing this idea, and show numerically that it captures qualitatively the relationships between elasticity, rigidity, thermodynamics and fragility. 
The thermodynamic properties of this model can be treated theoretically within a good accuracy in the temperature range we explore. Our key result is the following physical picture:  when there is an abundance
of soft elastic modes, elastic frustration  
vanishes, in the sense that 
a limited number of directions in phase space cost energy. Only those directions contribute to the specific heat, which is thus small. Away from the critical point, elastic frustration increases: 
more degrees of freedom contribute to the jump of specific heat, which increases while the boson peak is reduced.

\section{Model}

Our main assumption is 
that in a super-cooled liquid, nearby particles can organize themselves into a few distinct configurations. Consider for example covalent networks sketched in Fig.~1, where we use the label $\langle ij\rangle$ to indicate the existence of a covalent bond between particles $i$ and $j$. If two  covalent bonds $\langle 12\rangle$ and $\langle34\rangle$  are adjacent, there exists locally another configuration for which these bonds are broken, and where the bonds $\langle13\rangle$ and $\langle24\rangle$ are formed instead. These two configurations do not have the same energy in general. Moreover going from one configuration to the other generates a local strain, which creates an elastic stress  that propagates in space. In turn, this stress changes the energy difference between local configurations elsewhere in the system. This process leads to an effective interaction between local configurations at different locations. 

\begin{figure}
   \begin{center}   
   \resizebox{8.7cm}{!}{\includegraphics{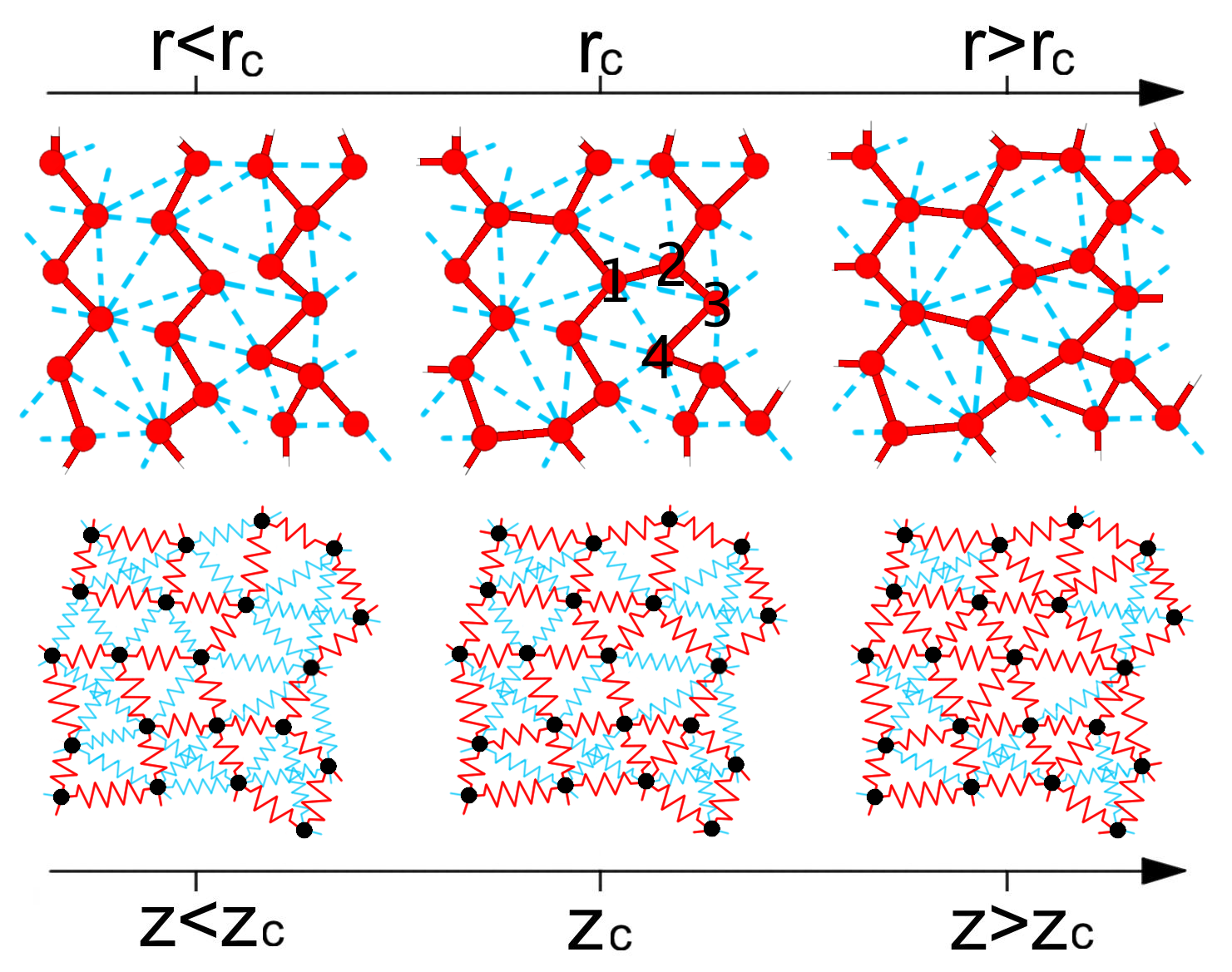}}     
   \caption{Top row:  sketches of covalent networks with different mean valence $r$ around the valence $r_c$: red solid lines represent covalent bonds; cyan dash lines represent van der Waals interactions. Bottom:  sketch of our elastic network model with varying coordination number $z$ (defined as the average number of strong springs in red) around Maxwell threshold $z_c$; cyan springs have a much weaker stiffness, and model weak interactions.}
  \label{fig1}
   \end{center}
\end{figure}

Our contention is that even a simple description of the  local configurations  -- in our case we will consider two-level systems, and we will make the approximation that the elastic properties do not depend on the levels -- can capture several unexplained aspects of super-cooled liquids, as long as the salient features of the elasticity of amorphous materials are taken into account. 
To incorporate in particular the presence of soft modes in the vibrational spectrum  we  consider random elastic networks. 
The elasticity of three types of networks have been studied extensively: networks of springs randomly deposited on a lattice \cite{GarbocziThor1985}, on-lattice self-organized networks \cite{Boolchand} and off-lattice random networks with small spatial fluctuations of coordination \cite{Wyartmaha,gustavo,wouter09}. 
We shall  consider the last class of networks, which  are known to 
capture correctly the scaling properties of elasticity near jamming, and can be treated analytically \cite{Wyart05,gustavo,wyart2010}.  It is straightforward  to extend our model to self-organized networks \footnote[1]  {If  the network is assumed to present a rigidity window  for which a subpart of the system is critical, we expect that within this window thermodynamics and fragility would behave as the critical point $z=z_c$ in the present model.}. 
 In our model two kinds of springs connect the $N$ nodes of the network: strong ones, of stiffness $k$ and coordination $z$, and weak ones,  of stiffness $k_{\rm w}$ and coordination $z_{\rm w}$.  These networks undergo a rigidity 
transition as $z$ crosses $z_c=2d$, where $d$ is the spatial dimension. For $z<z_c$ elastic stability is guaranteed by the presence of the weak springs. As indicated in Fig.~1, this situation is similar to covalent networks, where the weak Van der Waals interactions are required to insure stability when the valence $r$ is smaller than its critical value $r_c$.

Initially when our network is built, every spring $\langle ij\rangle$ is at rest: the rest length follows $l^0_{\langle ij\rangle}=||{\bf R}_i^0-{\bf R}_j^0||$, where ${\bf R}_i^0$ is the initial position of the node $i$. To allow for local changes of configurations we shall consider that any strong spring $\langle ij\rangle$ can switch between two rest lengths: $l_{\langle ij\rangle}=l^0_{\langle ij\rangle}+\epsilon \sigma_{\langle ij\rangle}$, where  $\sigma_{\langle ij\rangle}=\pm1$ is a spin variable. 
There are thus two types of variables: the $N_s\equiv zN/2$ spin variables $\{\sigma_{\langle ij\rangle}\}$, which we shall denote using the ket notation $|\sigma\rangle$, and the $Nd$ coordinates of the particles denoted by $|{\bf R}\rangle$. 
The elastic energy ${\cal E}( |{\bf R}\rangle,|\sigma\rangle)$ is a function of both types of variables.
The inherent structure energy  ${\cal \tilde H}(|\sigma\rangle)$ associated with any configuration $|\sigma\rangle$ is defined as:
\be
\label{2}
{\cal \tilde H}(|\sigma\rangle)\equiv \hbox{ min}_{|{\bf R}\rangle}{\cal E}(|{\bf R}\rangle,|\sigma\rangle)\equiv k\epsilon^2 {\cal H}(|\sigma\rangle) 
\ee
where we have introduced the dimensionless Hamiltonian ${\cal H}$. 
We shall consider the limit of small $\epsilon$, where the vibrational energy is simply that of harmonic oscillators. In this limit all the relevant information is contained in the inherent structures energy, since including the vibrational energy would increase the specific heat by a constant, which does not contribute to the jump of that quantity at the glass transition.
In this limit,  linear elasticity implies the form:
\be
\label{3}
{\cal H}(|\sigma\rangle)= \frac{1}{2} \sum_{\gamma, \beta} {\cal G}_{\gamma,\beta}\sigma_\gamma\sigma_\beta + o(\epsilon^2)\equiv \frac{1}{2}\langle \sigma|{\cal G}|\sigma\rangle +o(\epsilon^2)
\ee
where $\gamma$ and $\beta$ label strong springs, ${\cal G}_{\gamma,\beta}$ is the Green function describing how a dipole of force applied on the contact $\gamma$ changes the amplitude of the force in the contact $\beta$. 
Note that models  where some kind of defects interact elastically, leading to Hamiltonians similar in spirit to that of Eq.(\ref{3}), have been proposed to investigate the low-temperature properties of glasses \cite{sehtna} and super-cooled liquids \cite{lubchenko}. These models however assume continuum elasticity, unlike our model which can incorporate the effects of a rigidity transition and the presence of a boson peak.

${\cal G}_{\gamma,\beta}$ is computed in  Supplementary Information (SI)  part A and reads: 

\be
\label{green}
{\cal G}= {\cal I}-{\cal S}_{\rm s}{\cal M}^{-1}{\cal S}^t_{\rm s}
\ee
where ${\cal I}$ is the identity matrix, and   ${\cal S}_{\rm s}$ and the dimensionless stiffness matrix  ${\cal M}$ are standard linear operators connecting forces and displacements in elastic networks \cite{calladine}. They can be formally written as:
\begin{eqnarray}
\label{stiffM}
{\cal M}&=&{\cal S}^t_{\rm s}{\cal S}_{\rm s}+\frac{k_{\rm w}}{k}\,{\cal S}^t_{\rm w}{\cal S}_{\rm w},\\
{\cal S}_{\bullet}&=&\sum_{\langle ij\rangle_\bullet\equiv \gamma}| \gamma \rangle {\bf n}_{ij} (\langle i | -\langle j |)\nonumber
\end{eqnarray}
where $\langle i| {\bf R}\rangle\equiv {\bf R}_i$, $\langle ij\rangle_{\rm \bullet}$ indicate a summation over the strong springs ($\bullet=\rm s$) or the weak springs ($\bullet=\rm w$). 
 ${\cal S}_{\bullet}$ is a $N_\bullet\times dN$ matrix which projects any displacement field   onto the contact space of strong  or weak  springs. The components of this linear operator   are uniquely determined by  the unit vectors ${\bf n}_{ij}$   directed  along the contacts $\langle ij\rangle$ and point toward the node $i$.  

Finally note that the topology of the elastic network is frozen in our model. This addition of frozen disorder is obviously an approximation, as the topology itself should evolve as local configurations change. 
Our model thus misses the evolution of elasticity with temperature, that presumably affects the slowing down of the dynamics \cite{dyre} and gives a vibrational contribution to the jump of specific heat \cite{dyre,matPRL}. Building models which incorporate this possibility, while still  tractable numerically and  theoretically, remains a challenge. 

\section{Simulation}

\subsection{Network structure} Random networks with weak spatial fluctuations of coordination can be generated from random packings  of compressed soft particles \cite{Wyartmaha,gustavo,wouter09}. We consider packings with periodic boundary conditions. 
The centers of the particles correspond to the nodes of the network,  of unit mass $m=1$, and un-stretched springs of  stiffness $k=1$ 
 are put between particles in contact. 
 Then springs are removed,
preferably where the local coordination is high, so as to achieve the desired coordination $z$. 
In a second phase, $N_{\rm w}$ weak springs are added between the closest unconnected pairs of nodes.
The relative effect of those weak springs is best characterized by $\alpha\equiv (z_{\rm w}/d) (k_{\rm w}/k)$, which we modulate by fixing $z_{\rm w}=6$ and changing $(k_{\rm w}/k)$.
Note that an order of magnitude estimate of $\alpha$ in covalent glasses can be obtained by comparing  the behavior of the shear modulus $\ G$ in the elastic networks \cite{Wyartmaha} and in network glasses near the rigidity transition. As shown In Fig.~6 of  SI part B, this comparison yields the estimate that $\alpha \in [0.01,0.05]$.

\subsection{Thermodynamics} We introduce the rescaled temperature $T={\tilde T}k_B/(k\epsilon^2)$ where ${\tilde T}$ is the temperature. 
To equilibrate the system, we perform a one spin-flip Monte Carlo algorithm. The energy ${\cal H}$ of configurations are computed using Eq.(\ref{3}). 
We  use 5 networks of $N=256$ nodes in two dimensions and $N=216$ in three dimensions, each run  with 10 different initial configurations. Thus our results are averaged on these 50 realizations. We perform $10^9$ Monte Carlo steps  at each $T$.  The time-average inherent structure energy $E=\langle {\cal H}\rangle$ is calculated as a function of temperature, together with the specific heat $C_v=\partial E/\partial T$.  The intensive quantity  $c(T)\equiv C_v/N_s$ is represented in Fig.~2 for various excess coordination $\delta z=z-z_c$ and $\alpha=3\times10^{-4}$. We observe that the specific heat increases under cooling, until the glass transition temperature $T_g$ where $c(T)$ rapidly vanishes, indicating that the system falls out of equilibrium. 

\begin{figure}
   \begin{center}
      \vspace{-0.5cm}
     \rotatebox{+0}{\resizebox{8.7cm}{!}{\includegraphics{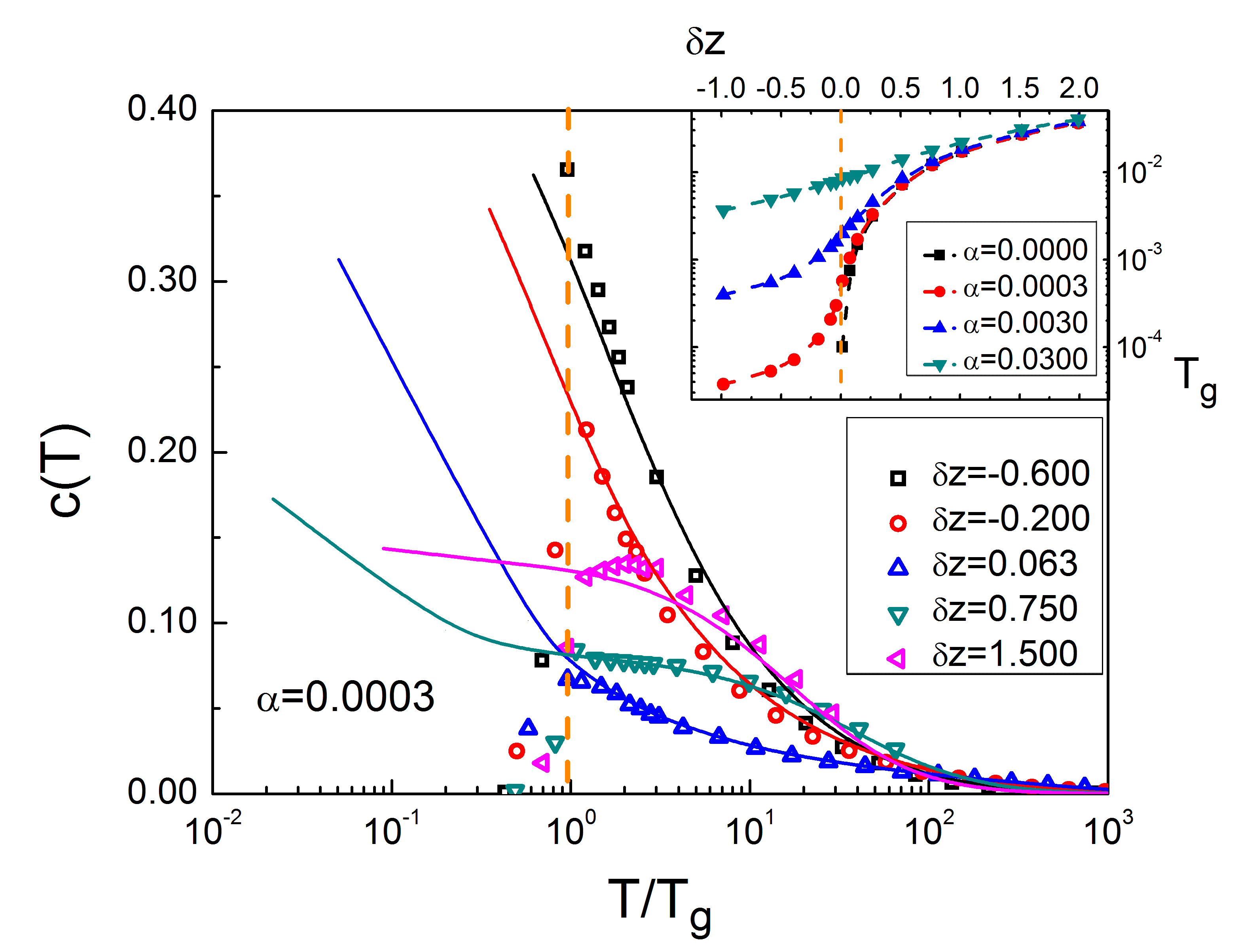}}}
     \caption{Specific heat $c(T)$ {\it v.s.} rescaled temperature $T/T_g$ for various excess coordination $\delta z\equiv z-z_c$ as indicated in legend, for $\alpha=3\times10^{-4}$ and $d=2$.  $c(T)$ displays a jump at the glass transition. Solid lines are theoretical predictions, deprived of any fitting parameters, of our mean-field approximation. They terminate at the Kautzman temperature $T_K$. Inset: glass transition temperature $T_g$ {\it vs} $\delta z$ for several amplitude of weak interactions $\alpha$, as indicated in legend. }
     \label{f2A}
   \end{center}
\end{figure}

\begin{figure}[b]
   \begin{center}
     \rotatebox{+0}{\resizebox{8.7cm}{!}{\includegraphics{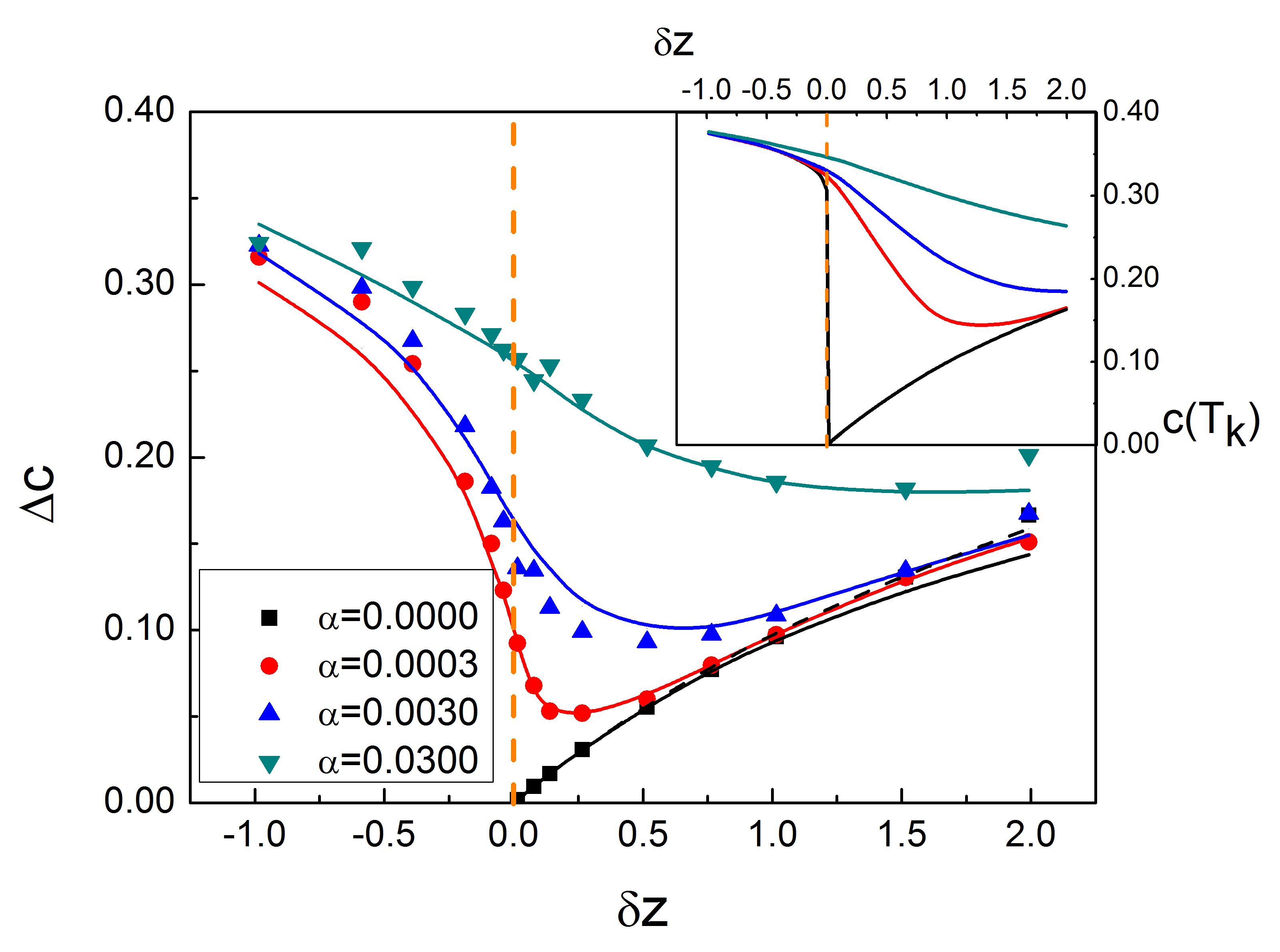}}}
     \caption{Jump of specific heat $\Delta c$ versus excess coordination $\delta z$ in $d=2$ for different strength of weak springs $\alpha$, as indicated in legend. Solid lines are mean-field predictions not enforcing the orthogonality of the $|\delta r_p\rangle$, dashed-line corresponds to the ROM where orthogonality is enforced. In both cases the specific heat is computed at the numerically obtained temperature $T_g$. Inset: theoretical predictions for $\Delta c$ {\it vs} $\delta z$ computed at the theoretical temperature $T_K$.}
     \label{f2B}
   \end{center}
\end{figure}

The amplitude of $c(T)$ just above $T_g$ thus corresponds to the jump of specific heat $\Delta c$ \footnote[2] {In our approach the absolute value of the specific heat will depend on the minimal number of particles needed to generate distinct local configurations, and on the number of configurations such a group can generate.  If those two numbers are of order one, our model predicts that  $\Delta c$ is of order one per particle or ``beads",  as observed near the glass transition \cite{woly}.}, and is shown in Fig.~3. Our key finding is that as the coordination increases,  $\Delta c(z)$ varies non-monotonically and is  minimal in the vicinity of the rigidity transition for all values of $\alpha$ investigated, as observed experimentally \cite{ANGELLCHALCO}. This behavior appears to result from a sharp asymmetric transition at $\alpha\to 0$. For $z>z_c$ we observe that $\Delta c(z)\propto \delta z$. The jump in specific heat thus vanishes as $\delta z\rightarrow 0^+$ where the system can be called  ``perfectly strong".  For $z<z_c$, $\Delta c$ is very rapidly of order one. 
When $\alpha$ increases, this sharp transition becomes a cross-over, marked by a minimum of $\Delta c(z)$ at some coordination larger but close to $z_c$.

\subsection{Dynamics} To characterize the dynamics  we compute the correlation function $C(t)=\frac{1}{N_s}\langle\sigma(t)|\sigma(0)\rangle$, which decays to zero at long time in the liquid phase. We define the relaxation time $\tau$ as $C(\tau)=1/2$, and the glass transition temperature $T_g$ as $\tau(T_g)/\tau(\infty)=10^5$. Finite size effects on $\tau$ appear to be weak, as shown in SI part C.
The Angell plot representing the dependence of $\tau$ with inverse rescaled temperature is shown in the inset of Fig.~4. It is found that the dynamics follows an Arrhenius behavior for $\alpha\rightarrow 0$ and $z\approx z_c$. Away from the rigidity transition, the slowing down of the dynamics is faster than Arrhenius. To quantify this effect we compute the fragility $m\equiv\frac{\partial\log_{10}\tau}{\partial(T_g/T)}|_{T=T_g}$, whose variation with coordination is presented in Fig.~4. Our key finding is that for all weak interaction amplitudes $\alpha$ studied, the fragility depends non-monotonically on coordination and is minimal near the rigidity transition, again as observed empirically \cite{angell2} in covalent liquids.
 As was the case for the thermodynamic properties, the fragility appears to be controlled by a critical point present at $\alpha=0$ and $z=z_c$ where the liquid is strong, and the dynamics is simply Arrhenius. As the coordination changes and $|\delta z|$ increases, the liquid becomes more fragile. The rapid change of fragility near the rigidity transition is smoothed over  when the amplitude of the weak interaction $\alpha$ is increased. 

\subsection{Correlating boson peak and fragility} The presence of soft elastic modes in glasses can be analyzed by considering the maximum of  $Z(\omega)\equiv \frac{D(\omega)}{D_D(\omega)}$ \cite{phillips_book}, where $D(\omega)$ is the vibrational density of states and $D_D(\omega)\propto \omega^2$ the Debye model for this quantity. The maximum  of $Z(\omega)$, $A_{BP}= Z(\omega_{BP})$ defines the boson peak frequency $\omega_{BP}$ \cite{phillips_book} and it normalized amplitude $A_{BP}$ \cite{novikov}. The inverse of $A_{BP}$ was shown to strongly correlate with  fragility \cite{novikov,Ngai} both in molecular liquids and covalent networks. 

To test if our model can capture this behavior we compute the density of states via a direct diagonalization of the stiffness matrix, see Eq.(\ref{stiffM}).  To compute $Z(\omega)$ the Debye density of states is estimated as $D(\omega)\sim \omega^2/G^{3/2}$ where $G$ is the shear modulus (bulk and shear moduli scale identically in this model, see e.g. \cite{wyart2010}). Then we extract the maximum $A_{BP}=Z(\omega_{BP})$. The dependence of $1/A_{BP}$ is represented in Fig.~5 and shows a minimum near the rigidity transition, and even a cusp in the limit $\alpha\rightarrow 0$.  
This behavior can be explained in terms of previous theoretical results on the density of states near the rigidity transition, that supports that $1/A_{BP}\sim |\delta z|^{1/2}$ when $\alpha\rightarrow 0$ \footnote[3]{ When $\alpha \rightarrow 0$ and $\delta z>0$, $\omega_{BP}\sim \delta z$ and $D(\omega_{BP})\sim 1 $ \cite{Wyart05}, whereas $G\sim \delta z$ \cite{Wyart053}, leading to $1/A_{BP}\sim \sqrt{\delta z}$. For $\delta z<0$, $G\sim -\alpha/\delta z$ \cite{Wyartmaha}. On the other hand the boson peak is governed by the fraction $\sim \delta  z$ of floppy modes, which gain a finite frequency $\sim \sqrt{\alpha}$ \cite{gustavo} thus we expect $D(\omega_{BP})\sim -\delta z/\sqrt{\alpha}$ and $1/A_{BP}\sim \sqrt{-\delta z}$.}.

Fig.~5 shows that $1/A_{BP}$ and the liquid fragility $m$ are correlated in our model, thus capturing observations in molecular liquids. The model also predicts  that $1/A_{BP}$ and the jump of specific heat are correlated. 
 Note that the correlation between fragility and $A_{BP}$ is not perfect, and that two branches, for glasses with low and with high coordinations, are clearly distinguishable. In general we expect physical properties  to depend on the full structure of the density of states, as will be made clear for the thermodynamics of our model below. The variable $A_{BP}$, which is a single number, cannot capture fully this relationship. In our framework it is a useful quantity however, as it characterizes well the proximity of the jamming transition. 

\section{Theory}
\subsection{Thermodynamics in the absence of weak interactions ($\alpha=0$)} 
In the absence of weak springs the thermodynamics is non-trivial if $z\geq z_c$, otherwise the inherent structure energies are all zero. Then Eq.(\ref{stiffM}) implies ${\cal M}={\cal S}^t_{\rm s} {\cal S}_{\rm s}$, and Eq.(\ref{green}) leads to ${\cal G}={\cal I}-{\cal S}_{\rm s}({\cal S}^t_{\rm s}{\cal S}_{\rm s})^{-1}{\cal S}^t_{\rm s}$. Inspection of this expression indicates that ${\cal G}$ is  a projector on the kernel of ${\cal S}^t_{\rm s}$, 
which is generically of dimension $N_s-Nd\equiv \delta z N/2$. This kernel corresponds to all the sets of contact forces that balance forces on each node \cite{Wyart053}.  We denote by $|\delta r_p\rangle, p=1,...,\delta z N/2$ an orthonormal basis of this space. We may then rewrite ${\cal G}=\sum_p |\delta r_p\rangle\langle \delta r_p|$ and Eq.(\ref{3}) as:
\be
\label{7}
{\cal H}(|\sigma\rangle)=\frac{1}{2}\sum_{p=1}^{\delta zN/2} \langle \sigma|\delta r_p\rangle^2
\ee
Eq.(\ref{7}) is a key result, as it implies that near the rigidity transition the number $\delta z N/2$ of directions of phase space that cost energy vanishes. Only those directions can contribute to the specific heat, which must thus vanish linearly in $\delta z$ as the rigidity transition is approached from above.

Eq.(\ref{7}) also makes a connection between strong liquids in our framework and well-know problems in statistical mechanics. In particular Eq.(\ref{7}) is similar to that describing Hopfield nets \cite{Hopfield} used to store $\delta z N/2$ memories consisting of the spin states $|\delta r_p\rangle$. The key difference is the sign: in Hopfield nets memories correspond to  meta-stables states, whereas in our model the vectors $|\delta r_p\rangle$ corresponds to maxima of the energy. A particularly interesting case is $\delta z N/2=1$, the closest point to the jamming transition which is non-trivial.  In this situation  the sum in Eq.(\ref{7}) contains only one term: ${\cal H}(|\sigma\rangle)= \frac{1}{2} \langle \sigma|\delta r_1\rangle^2= \frac{1}{2} (\sum_{\alpha=1}^{N_s} \delta r_{1,\alpha} \sigma_\alpha)^2$. 
This Hamiltonian corresponds to the  NP complete partitioning problem \cite{hayes}, where given a list of numbers (the $\delta r_{1,\alpha}$) one must partition this list into two groups 
whose sums are as identical as possible. Thermodynamically this problem is known \cite{mertens} to map into the random energy model \cite{REM} where energy levels are randomly distributed. 
 
\begin{figure}
   \begin{center}
   \vspace{-0.4cm}
     \rotatebox{+0}{\resizebox{8.7cm}{!}{\includegraphics{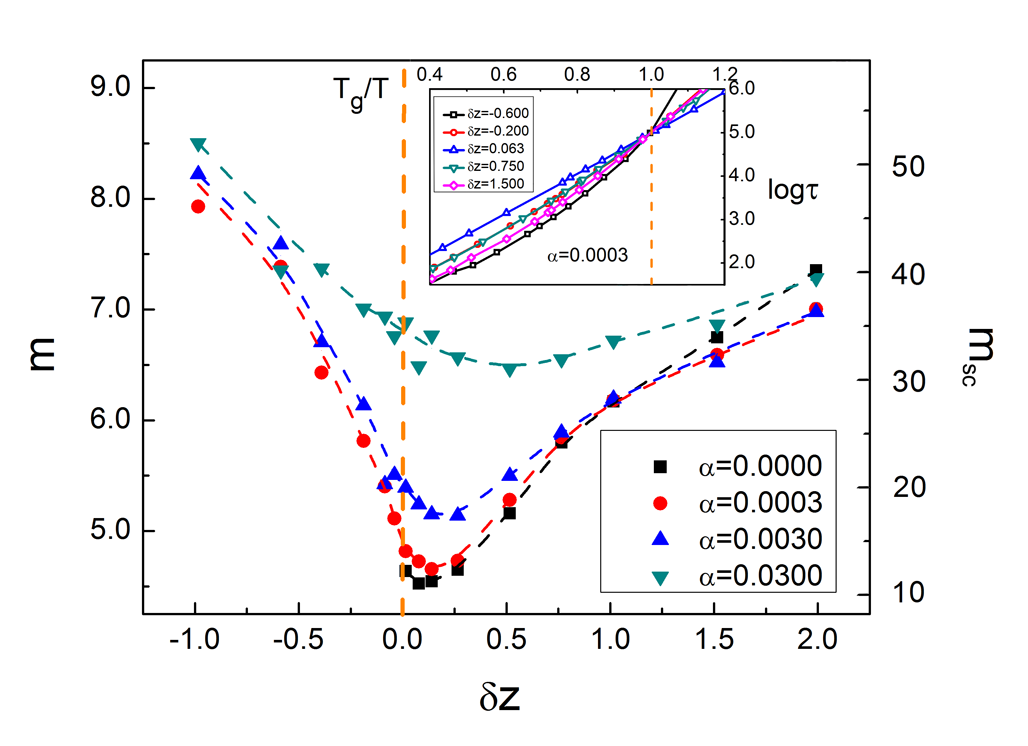}}}
        \vspace{-0cm}
     \caption{Fragility $m$ and  rescaled fragility $m_{sc}$ extrapolated to the experimental dynamical range (see SI part D for a definition) versus excess coordination $\delta z$ for different strength of weak interactions $\alpha$ as indicated in legend, in $d=2$. Dash dot lines are guide to the eyes, and reveal the non-monotonic behavior of $m$ near the rigidity transition.   Inset: Angell plot representing $\log\tau$ {\it v.s.} inverse temperature $T_g/T$ for different $\delta z$ and $\alpha=3\times10^{-4}$. }
     \label{f3}
   \end{center}
\end{figure}

\begin{figure}
   \begin{center}
     \rotatebox{+0}{\resizebox{8.7cm}{!}{\includegraphics{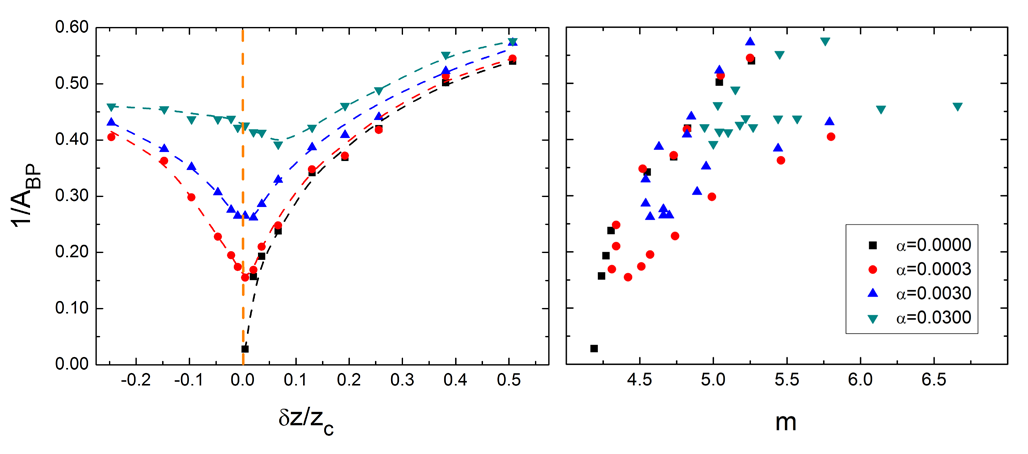}}}
     \caption{Left:  Inverse boson peak  amplitude $1/A_{BP}$ versus excess coordination $\delta z$ in our $d=3$ elastic network model, for different weak interaction strenghts as indicated in legend. Dash dot lines are drawn to guide one's eyes. Right: Inverse boson peak  amplitude  $1/A_{BP}$ versus fragility $m$ for different weak springs $\alpha$.}
     \label{f4}
   \end{center}
\end{figure}

It is in general very difficult  to compute the thermodynamic functions
of the problem defined by Eq.(\ref{7})  because the vectors $|\delta r_p\rangle$ present spatial correlations,
as must be the case since the amplitude of the interaction kernel ${\cal G}_{\gamma,\beta}$ must decay with distance. However this effect is expected to be mild near the rigidity transition. Indeed there exists a diverging  length scale at the transition, see  \cite{gustavo} for a recent discussion, below which ${\cal G}_{\gamma,\beta}$ is dominated by fluctuations and decays mildly with distance. Beyond this length scale ${\cal G}_{\gamma,\beta}$ presents a dipolar structure, as in a standard continuous elastic medium.  We shall thus assume that $|\delta r_p\rangle$ are random unitary  vectors, an approximation of mean-field character expected to be good near the rigidity transition.

 Within this approximation, the thermodynamic properties can be derived for any spectrum of ${\cal G}$ \cite{ROM}. If the orthogonality of the vectors $|\delta r_p\rangle$ is preserved,  the Hamiltonian of Eq.(\ref{7}) corresponds to the Random Orthogonal Model (ROM) whose thermodynamic properties have been derived \cite{ROM} as well as some aspects of the dynamics \cite{ROmodels}. Comparison of the specific heat of our model and the ROM predictions of \cite{ROM} is shown in Fig.~3 and are found to be very similar. For sake of simplicity, in what follows we shall also relax  the orthogonal condition on the vectors $|\delta r_p\rangle$. This approximation allows for a straightforward analytical treatment in the general case $\alpha\neq 0$, and is also very accurate near the rigidity transition since the number of vectors $\delta z N/2$  is significantly smaller than the dimension of the space $dN$ they live in, making random vectors effectively orthogonal. Under these assumptions we recover the random Hopfield model with negative temperature.

In the parameter range of interest, the Hopfield free energy ${\cal F}={\overline{\ln({\cal Z})}}$ (here $\overline{(...)}$ represents the disorder average on the $|\delta r_p\rangle$) is approximated very precisely by the annealed free energy $\ln({\overline {\cal Z}})$  (this is obviously true for the number partitioning problem that maps to the Random Energy Model), which can be easily calculated. Indeed in our approximations the quantities $X_p\equiv\langle \sigma|\delta r_p\rangle$ are independent gaussian random variables  of variance one, and:
\begin{equation}
\label{8}
{\cal {\overline Z}}\propto \int\left(\prod_{p=1}^{N\delta z/2}dX_p\frac{1}{\sqrt{2\pi}}e^{-X_p^2/2}\right)e^{-\frac{1}{2T}\sum_pX_p^2}
\end{equation}
Performing the Gaussian integrals we find:
\be
\label{9} c(T)=\frac{\delta z}{2z}\frac{1}{(1+T)^2}
\ee
The Kautzman temperature defined as $s(T_K)=0$ is found to follow $T_K\approx\frac{2}{e}2^{-2z/\delta z}$.
 Eq.(\ref{9}) evaluated at $T_g$ is tested against the numerics in Fig.~3 and performs remarkably well for the range of coordination probed. 

\subsection{General case ($\alpha\neq 0$)} 
To solve our model analytically in the presence of weak interactions, we make the additional approximation that the associated coordination $z_{\rm w}\rightarrow \infty$, while keeping $\alpha\equiv z_{\rm w} k_{\rm w}/(kd)$ constant. 
In this limit weak springs lead to an effective interaction between each node and  the center of mass of the system, that is motionless. Thus the restoring force stemming from  weak interactions $|{\bf F}_{\rm w}\rangle$ follows $|{\bf F}_{\rm w}\rangle=-\alpha |\delta{\bf R}\rangle$, leading to a simple expression in  the stiffness matrix Eq.(\ref{stiffM})  for the weak spring contribution $\frac{k_{\rm w}}{k}{\cal S}_{\rm w}^t{\cal S}_{\rm w}=\alpha  {\cal I} $. It is useful to perform the eigenvalue decomposition:
\be
\label{99}
{\cal S}_{\rm s}^t{\cal S}_{\rm s}=\sum_\omega \omega^2 |\delta {\bf R}_\omega\rangle\langle\delta {\bf R}_\omega|
\ee
where $|\delta {\bf R}_\omega\rangle$ is the vibrational mode of frequency $\omega$ in the elastic network without weak interactions. We introduce the orthonormal eigenvectors in contact space $|\delta r_\omega\rangle\equiv {\cal S}_{\rm s} |\delta {\bf R}_\omega\rangle/\omega$ defined for $\omega>0$. For $\delta z<0$ these vectors form a complete basis of that space, of dimension $N_s$. When $\delta z>0$ however, this set is of dimension $Nd<N_s$, and it  must be completed by the kernel of ${\cal S}^t_{\rm s}$, i.e. the set of the $|\delta r_p\rangle, p=1,...,\delta z N/2$ previously introduced. Using this decomposition in Eq.(\ref{green},\ref{stiffM}) we find:
\be
\label{10}
{\cal H}(|\sigma\rangle)=\frac{1}{2}\sum_{p=1}^{\delta zN/2} \langle \sigma|\delta r_p\rangle^2+\frac{1}{2}\sum_{\omega>0} \frac{\alpha}{\alpha+\omega^2}\langle \sigma|\delta r_\omega\rangle^2
\ee
where the first term exists only for $\delta z>0$. Using the mean field approximation that the set of  $|\delta r_p\rangle$ and $|\delta r_\omega\rangle$ are random gaussian vectors,
the annealed free energy is readily computed, as shown in SI part E. We find in particular for the specific heat: 
\begin{equation}
\label{11}
c(T)=\frac{\delta z }{2z}\frac{\theta(\delta z)}{(1+T)^2}+ \frac{1}{2 N_s}\sum_{\omega>0}\left(\frac{\alpha}{\alpha+(\omega^2+\alpha)T}\right)^2
\end{equation}
where $\theta(x)$ is the unitary step function. To compare this prediction with our numerics without fitting parameters, we compute numerically the vibrational frequencies for each value of the coordination.
Our results  are again in excellent agreement with our observations, as appears in Figs.~\ref{f2A},~\ref{f2B}.

To obtain the asymptotic behavior near jamming, we  replace  the summation over frequencies in Eq.(\ref{11})  by an integral $\sum_{\omega>0}\rightarrow N_s\int \rm{d}\omega D(\omega)$. 
The associated density of vibrational modes $D(\omega)$ in such networks has been computed theoretically \cite{wyart2010,gustavo,Wyart05}. 
These results allows us to compute the scaling behavior of thermodynamic properties near the rigidity transition, see SI part F. We find that the specific heat increases monotonically with decreasing temperature.  Its value at the Kautzman temperature thus yields an  upper
bound on the jump of specific heat. In the limit $\alpha\rightarrow 0$, we find that a sudden discontinuity of the jump of specific heat occurs at the rigidity transition:
\ba
\label{12}c(T_K)&\sim& \frac{\delta z}{2z} \ \ \ \ \  \hbox{    for  }  \delta z>0\\
\label{13}\lim_{\delta z\rightarrow 0^-} c(T_K)&\sim&\frac{\pi z_c}{8z} \ \ \ 
\ea
Eq.(\ref{12}) states that adding weak interactions is not a singular perturbation for $\delta z>0$,  and we recover Eq.(\ref{9}). 
On the other hand for $\delta z<0$, the energy of inherent structures is zero in the absence of weak springs, which thus have a singular effect.  The relevant scale of temperature is then a function of $\alpha$. In particular we find that the Kautzman temperature is sufficiently low that all the terms in the second sum of Eq.(\ref{11}) contribute significantly to the specific heat, which is therefore large  as Eq.(\ref{13}) implies.
Thus as the coordination decreases below the rigidity transition, one goes discontinuously from a regime where at the relevant temperature scale the energy landscape consists of a vanishing number of costly directions in phase space, whose cost is governed by the strong interaction $k$, to a regime where the weak interaction $\alpha$ is the relevant one, and where at the relevant temperature scale all directions in phase space contribute to the specific heat.

Note that although the sharp change of thermodynamic behavior that occurs at the rigidity transition is important conceptually, empirically a smooth cross-over will always be observed. This is the case because (i) $\alpha$ is small but finite. As $\alpha$ increases this sharp discontinuity is replaced by a cross-over at a coordination $\delta z\sim \ln(1/\alpha)^{-1}$ (see SI part F) where $c(T_K,z)$ is minimal, as indicated in the inset of Fig.~3. (ii) The Kautzman temperature range is not accessible dynamically, i.e. $T_g>>T_K$ near the rigidity transition. Comparing Fig.~3 with its inset, our theory predicts that the minimum of $c(T_g)$ is closer to $z_c$ and more pronounced than at $T_K$.

\section{Discussion}
 
Previous work \cite{revue} has shown that well-coordinated glasses must have a small boson peak, which increases as the coordination (or valence for network glasses) is decreased toward the jamming (or rigidity) transition.
Here we have argued that as this process occurs,  elastic frustration vanishes:  thanks to the abundance of soft modes, any configuration (conceived here as a set of local arrangements of the particles) can relax more and more of its energy as jamming is approached from above. As a result, the effective number of degrees of freedom that cost energy and contribute to the jump of specific heat at the glass transition vanishes. 
As the coordination is decreased further below the rigidity transition, the scale of energy becomes governed by the weak interactions (such as Van der Waals) responsible for the finite elasticity of the glass. At that scale, all direction in phase space have a significant cost and the specific heat increases. This view potentially explains why linear elasticity strongly correlates to key aspects of the energy landscape in network and molecular glasses \cite{ANGELLCHALCO,angell,novikov,Ngai}. 
This connection we propose between structure and dynamics can also be tested  numerically. Our results are consistent with the observation that soft particles become more fragile when compressed away from the random close packing \cite{berthier2} \footnote[4]{ A quantitative treatment of the vicinity of random close packing would require  including the dependence of elasticity on temperature neglected here.}.  Another interesting parameter to manipulate is the amplitude of weak interactions, which can be increased  by adding long-range forces to the interaction potential \cite{Wyart053,Xu07}. According to our analysis, doing so should increase fragility, in agreement with existing observations \cite{berthier}. 

The model of the glass transition we introduced turns out to be a spin glass model,  with the specificity that (i) the interaction is dipolar in the far field, and that  (ii) the sign of the interaction is approximatively  random below some length scale $l_c$ that diverges near jamming, where the coupling matrix has a vanishingly small rank. Applying spin glass models to structural glasses have a long history. In particular the Random First Order Transition theory (RFOT) \cite{woly} is based on mean-field spin glass models that display a thermodynamic transition at some $T_K$  where the entropy vanishes.
A phenomenological description of relaxation in liquids near $T_K$  based on the nucleation of random configurations leads to a diverging time scale and length scale $\xi$ at $T_K$ \cite{woly,BB}. One limitation of this approach is that no finite dimensional spin models with two-body interactions have been shown to follow this scenario so far \cite{giuliocondmat}, and it would thus be important to know if our model does display a critical point at finite temperature.  Our model  will also allow one to investigate the generally neglected role of the action at a distance allowed by elasticity, characterized by a scale $l_c$. In super-cooled liquids heterogeneities of elasticity (that correlates to irreversible rearrangements) can be rather extended \cite{harrowell} suggesting that $l_c$  is large. This length scale may thus play an important role in a description of relaxation in liquids, and in deciphering the relationship between elastic and dynamical heterogeneities.

\begin{acknowledgments} We thank A. Grosberg, P. Hohenberg, E. Lerner, D. Levine, D. Pine, E. Vanden-Eijnden, M. Vucelja., A. Lef\`evre for discussions, and E.Lerner for discussions leading to Eq.(SI-1). This work has been supported primarily by the  National Science Foundation  CBET-1236378, and partially by the Sloan Fellowship, the NSF DMR-1105387, and the Petroleum Research Fund 52031-DNI9.
\end{acknowledgments}

\section{SUPPLEMENTARY INFORMATION}

\section{A. Stiffness and Coupling matrices}
Consider  a  network of $N$ nodes connected by $N_c$ springs. 
If an infinitesimal displacement field $|\delta {\bf R}\rangle $ is imposed on the nodes, the change of length of the springs  can be written as a vector $|\delta r\rangle$ of dimension $N_c$. For small displacements this relation is approximately linear: $|\delta r\rangle ={\cal S} |\delta {\bf R}\rangle $, where ${\cal S}$ is an $N_c\times Nd$ matrix.  To simplify the notation, we write ${\cal  S}$ as an $N_c\times N$ matrix of components of dimensions $d$, which gives ${\cal  S}_{\gamma,i}\equiv \partial r_\gamma/\partial {\bf R}_i=\delta_{\gamma,i} {\bf n}_\gamma$, where $\delta_{\gamma,i}$ is non-zero only if the contact $\gamma$ includes the particle $i$, and  ${\bf n}_\gamma$ is the unit vector in the direction of the contact $\gamma$, pointing toward the node $i$.  Using the bra-ket notation, we can rewrite   ${\cal  S}=\sum_{\langle ij\rangle\equiv \gamma}| \gamma \rangle {\bf n}_{\gamma} (\langle i | -\langle j |)\nonumber$, where the sum  is over all the springs of the network.  Note that the transpose ${\cal S}^t$ of ${\cal  S}$ relates the set of contact forces $|f\rangle$  to the set $|{\bf F}\rangle$ of unbalanced forces on the nodes:  $| {\bf F}\rangle={\cal  S}^t |f\rangle$, which simply follows from the fact that ${\bf F}_i=\sum_{\gamma} \delta _{\gamma,i} f_\gamma {\bf n}_\gamma=\sum_{\gamma} f_\gamma {\cal S}_{\gamma,i}$ \cite{calladine}. 

The stiffness matrix ${\cal \tilde M}$ is a linear operator connecting external forces to the displacements: ${\cal \tilde M}|\delta {\bf R}\rangle=|{\bf F}\rangle$. Introducing the $N_c\times N_c$ diagonal matrix ${\cal K}$, whose components are the spring stiffnesses ${\cal K}_{\gamma\gamma}=k_{\gamma}$, we have for harmonic springs $|f\rangle={\cal K} |\delta r\rangle$. Applying ${\cal  S}^t$ on each side of this equation, we get $|{\bf F}\rangle={\cal  S}^t |f\rangle={\cal  S}^t 
{\cal K} {\cal S} |\delta {\bf R}\rangle$, which thus implies \cite{calladine}:
\be
\label{4}
{\cal \tilde M}={\cal  S}^t {\cal K}{\cal  S}.
\nonumber
\ee
Let us assume that starting from a configuration where all springs are at rest, the rest lengths of the springs are changed by some amount $|y\rangle$. This will generate an unbalanced force  field $|{\bf F}\rangle ={\cal  S}^t{\cal K}|y\rangle$ on the nodes, leading to a displacement $|\delta {\bf R}\rangle={\cal \tilde M}^{-1}{\cal  S}^t{\cal K}|y\rangle$. The elastic energy ${\cal E}=\frac{1}{2}\langle y-\delta r|\mathcal{K}|y-\delta r\rangle$ is minimal for this displacement and the corresponding energy ${\cal \tilde H}$ is:
\be
\label{5}
{\cal \tilde H}(|y\rangle)=\frac{1}{2}\langle y|{\cal K}-{\cal K}{\cal  S}{\cal \tilde M}^{-1}{\cal  S}^t{\cal K}|y\rangle.
\tag{SI-1}
\ee
In our model,  $y_\gamma=0$ for weak springs and $y_\gamma=\epsilon \sigma_\gamma$ for strong springs of stiffness $k$, implying that ${\cal K} |y\rangle=k |y\rangle$. Introducing the dimensionless stiffness matrix ${\cal M}\equiv {\cal \tilde M}/k$ and the restriction ${\cal  S}^t_{\rm s}$ of the operator ${\cal  S}^t$ on the subspace of strong contacts of dimension $N_s$, i.e. ${\cal S}^t_{\rm s} |\sigma\rangle\equiv {\cal  S}^t|y\rangle$, Eq.(\ref{5}) yields:
\be
\label{6}
{\cal H}(|\sigma\rangle)=\frac{1}{2}\langle\sigma|{\cal G}|\sigma\rangle \hbox{ where } {\cal G}={\cal I}-{\cal S}_{\rm s}{\cal M}^{-1}{\cal S}^t_{\rm s},
\nonumber
\ee
where ${\cal I}$ is the identity matrix, and ${\cal G}$ is the coupling matrix used in the main text. Note that in our model the diagonal matrix ${\cal K}$ contains only two types of coefficients $k_{\rm w}$ and $k$, corresponding to the stiffnesses of weak springs and stiff springs respectively. Then the dimensionless  stiffness matrix can be written as ${\cal M}={\cal S}^t_{\rm s}{\cal S}_{\rm s}+\frac{k_{\rm w}}{k}\,{\cal S}^t_{\rm w}{\cal S}_{\rm w}$, where ${\cal S}^t_{\rm w}$ is the projection of the operator ${\cal S}^t$ on the subspace of weak contacts.
\begin{figure}[b]
   \begin{center}
   \resizebox{8.7cm}{!}{\includegraphics{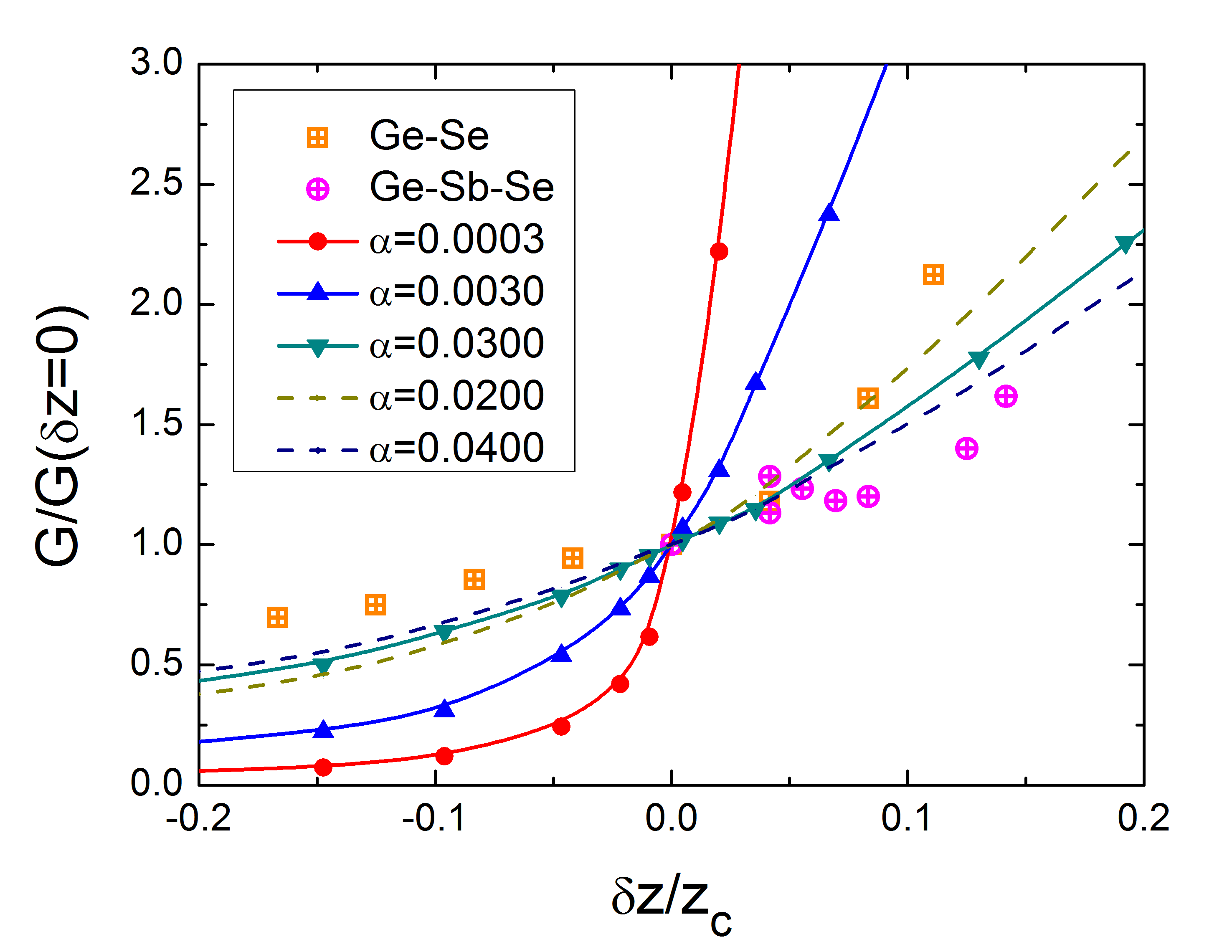}}
   \caption{Squares show the shear modulus $G$ normalized by its value at the rigidity threshold for Ge-Se, taken from Ref.~\cite{GeSe}. Circles show $G$ for Ge-Sb-Se, taken from Ref.~\cite{GeSbSe}. Lines display the shear modulus $G$ for network models in $d=3$  using  different $\alpha$, as indicated in the legend.}
     \label{A1}
   \end{center}
\end{figure}
\section{B. Shear Modulus of the random elastic networks}
An explicit  expression for the shear modulus ${G}$  of an elastic network can be found using linear response theory \cite{Lutsko,lerner}. In particular, let us consider  the shear on the $(x, y)$ plane.  In the contact vector space, a shear strain can be written as  $\delta|y^{sh}\rangle$ with $\delta\ll1$ which represents the amplitude of the strain and $|y^{sh}\rangle$ corresponds to a unit shear strain. The components of  $|y^{sh}\rangle$  are given by  $y^{sh}_\gamma=\frac{\Delta x_\gamma\Delta y_\gamma}{l_\gamma}$, where  $l_{\gamma}$ is the rest length of the spring $\gamma$, and  $\Delta x_{\gamma}$ and $\Delta y_{\gamma}$ are its projections along the x and y directions. From the last section (A), we can obtain  the total energy induced by a shear strain  ${\cal \tilde H}(|y^{sh}\rangle)\delta^2$; hence the shear modulus $G=2 {\cal \tilde H}(|y^{sh}\rangle)/ V$.

To estimate the value of $\alpha$, we consider the dependence of the shear modulus with coordination or valence in the vicinity of the rigidity transition, which is smooth for large $\alpha$ and sudden for small $\alpha$ in our networks, see Fig.~6.  Comparing networks and  real chalcogenide glasses we find that the cross-over in the elastic modulus is   qualitatively reproduced for  $\alpha\in [0.01,0.05]$.

\section{C.  Finite size effects on fragility}

To estimate the role of finite size effects on the dynamics, we use  two different system sizes $N=64$ and $N=256$. As shown in Fig.~7, the Angell plot for the relaxation time, and therefore our estimation of the fragility,  appears to be nearly independent of the system size. Note, however, that the correlation function $C(t)$  shows some finite size effects very close to the isostatic point $(z=z_c,\alpha=0)$, but that it does not affect our measure of $\tau$ significantly. 
In particular we find that near isostaticity, the distribution of relaxation time is broad for small systems, and becomes less and less so when the system size increases. 
We noticed that this effect also disappears if a two-spin flips Monte-Carlo is used, instead of the one-spin flip algorithm we perform. 

\section{D. Rescale fragility with different dynamical range}

The value of fragility depends on the definition of glass transition, in particular on the dynamical range. In super-cooled liquids the glass transition occurs when the relaxation time is about $10^{16}$ larger than the relaxation time at high temperature. Thus the dynamical range in experiments (which corresponds to the fragility of a perfectly Arrhenius liquid)  is $R=16$. In our simulation, the same quantity is $R=5$. It is possible however to rescale our values of fragility  to compare with experimental data, if we extrapolate the dynamics.
We shall assume a Vogel-Fulcher-Tammann (VFT)  relation at low temperature,
\[
\log_{10}\frac{\tau(T)}{\tau_0}=\frac{A}{T-T_0},
\]
We define the dynamical range as:
\[
R=\log_{10}\frac{\tau(T_g)}{\tau_0}=\frac{A}{T_g-T_0}.
\]
Thus we can express the fragility as: 
\be
\label{111}
m_R=\left.\frac{\partial\log_{10}\tau(T)/\tau_0}{\partial{T_g/T}}\right|_{T=T_g}=R+R^2\frac{T_0}{A}.
\tag{SI-2}
\ee
$T_0$ and $A$ are assumed to be independent of dynamical range.
Using the notation $m=m_5$ and $m_{sc}=m_{16}$ we get from Eq.(\ref{111}):
\[
m_{sc}=16+\frac{16^2}{5^{2}}(m-5)=10.24m-35.2
\]
The amplitude of fragility we find turn out to be comparable to experiments when $\alpha=0.03$, in particular for $z
\geq z_c$. For the smallest coordination explored our results underestimate somewhat the fragility, slightly above 50 in our model and about 80 experimentally.
This is not surprising considering that our model is phenomenological, and the extrapolation we made to compare different dynamical ranges.

\section{E. Canonical ensemble approach with weak springs}

In the case where $\alpha\neq0$, the annealed free energy can be easily calculated under the assumption that  $|\delta r_p\rangle$ and $|\delta r_{\omega}\rangle$ are random Gaussian vectors. The Hamiltonian in Eq.(10) can be rewritten as:
$$ {\cal H}=\frac{1}{2}\sum_{p=1...\delta zN/2} X_p^2+\frac{1}{2}\sum_{\omega>0} \frac{\alpha}{\alpha+\omega^2}X_\omega^2,$$
where  $X_p=\langle\delta r_{p}|\sigma\rangle$ and $X_{\omega}=\langle\delta r_{\omega}|\sigma\rangle$ represent  independent  random variables for each configuration $|\sigma\rangle$.  In the thermodynamic limit the random variables $X_p$ and $X_\omega$ are Gaussian distributed  with zero mean and unit variance. The averaged partition function is  given by: 

\be
\label{a1}
\begin{split}
\overline{\mathcal{Z}}&=2^{N_s}\int e^{-{\cal H}/T}\,\prod_{p}\frac{e^{-\frac{X_{p}^2}{2}}}{\sqrt{2\pi}}\mathrm{d}X_{p}\prod_{\omega}\frac{e^{-\frac{X_{\omega}^2}{2}}}{\sqrt{2\pi}}\mathrm{d}X_{\omega}\\
&=2^{N_s}\prod_{p=1}^{\delta zN/2}\left(1+\frac{1}{T}\right)^{-1/2}\prod_{\omega>0}\left(1+\frac{\alpha/T}{\alpha+\omega^2}\right)^{-1/2}.
\end{split}
\nonumber
\ee

\begin{figure}[t]
  \label{BB}
   \begin{center}
   \resizebox{8.7cm}{!}{\includegraphics{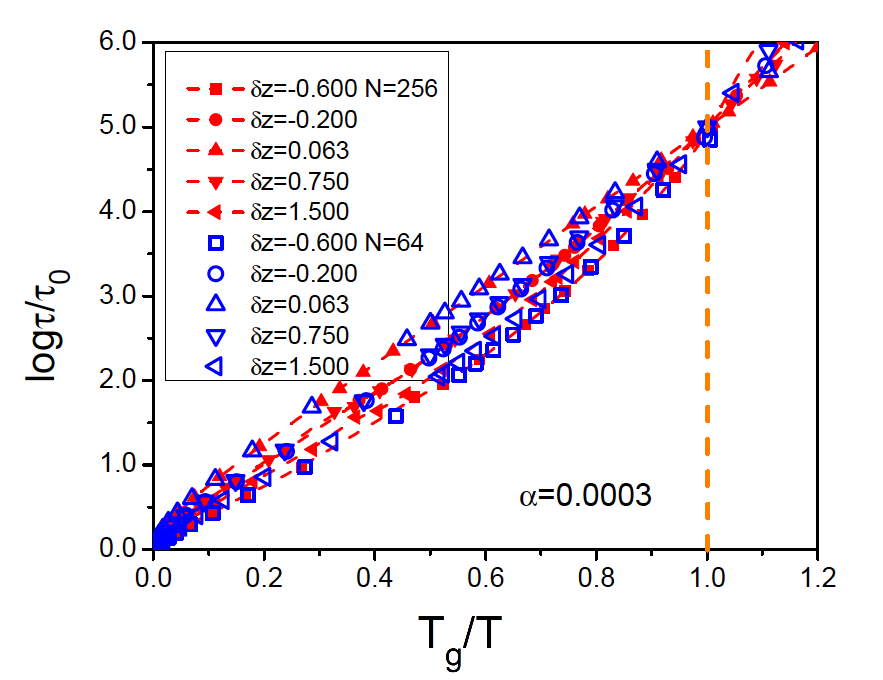}}
   \caption{Angell plot representing $\log \tau$ {\it v.s.} inverse temperature $T_g/T$ for different $\delta z$ and two system sizes $N=64$ and $N=256$, $\alpha=0.0003$.}
      \end{center}
\end{figure}

From the average partition function the density of free energy per spring $f(T)$  and any other thermodynamic quantities are readily computed. In particular,  the energy density $ \varepsilon(T)$, the specific heat $c(T)$ and the entropy density $s(T)$ write:
 
\begin{equation}
\label{a2}
f(T)=\frac{T}{2N_s}\left[\sum_{p=1}^{\delta zN/2}\ln\left(1+\frac{1}{T}\right)+\sum_{\omega>0}\ln\left(1+\frac{\alpha/T}{\alpha+\omega^2}\right)\right]-T\ln2
\nonumber 
\end{equation}
\begin{equation}
\label{a3}
\varepsilon(T)=\frac{1}{2N_s}\left[\sum_{p=1}^{\delta zN/2}\frac{T}{1+T}+\sum_{\omega>0}\frac{\alpha T}{\alpha+(\omega^2+\alpha)T}\right]
\nonumber
\end{equation}
\begin{equation}
\label{a4}
c(T)=\frac{1}{2N_s}\left[\sum_{p=1}^{\delta zN/2}\frac{1}{\left(1+T\right)^2}+\sum_{\omega>0}\left(\frac{\alpha}{\alpha+(\omega^2+\alpha)T}\right)^2\right]
\tag{SI-2}
\end{equation}
\begin{equation}
\label{a5}
\begin{split}
s(T)=&\ln2-\frac{1}{2N_s}\sum_{p=1}^{\delta zN/2}\left[\ln(1+\frac{1}{T})-\frac{1}{1+T}\right]\\
&-\frac{1}{2N_s}\sum_{\omega>0}\left[\ln\left(1+\frac{\alpha/T}{\alpha+\omega^2}\right)-\frac{\alpha}{\alpha+(\omega^2+\alpha)T}\right].
\end{split}
\nonumber
\end{equation}

In the limit $\alpha\to0$, for any finite temperature $T$, the  sum over the  vibration modes ($\omega>0$) vanishes, and we recover the expressions in the  absence of weak springs  for pure rigid networks. Note that  Eq.(\ref{a4}) corresponds to Eq.(11) in the article.

\section{F.  Continuous density of states limit: Analytical results}

In the thermodynamic limit $N\to\infty$, we can replace the sum over frequencies by an integral: $\sum_{\omega>0}\to N_s\int\mathrm{d}\omega D(\omega)$ for $\delta z\leq0$, and $\sum_{\omega>0}\to Nd\int\mathrm{d}\omega D(\omega)$ for $\delta z>0$. The density of states $D(\omega)$ is the distribution of vibrational modes of random elastic networks,  which    has been  computed theoretically~\cite{Wyart05,gustavo,wyart2010}.
There are two frequency scales in the random network : $\omega^*\sim|\delta z|$  above which a plateau of soft modes exist,  and a cut-off  frequency $\omega_c\sim1$. Below $\omega^*$, rigid networks show plane wave modes \cite{Wyartmaha,Wyart05,wyart2010} with a characteristic Debye regime $D(\omega)\sim \omega^2$, unlike  floppy networks, which show no modes in this gap \cite{gustavo}. 

It turns out that the Debye  regime contribution to the integrals is negligible near the jamming threshold.
To capture the scaling behavior near jamming, we approximate  $D(\omega)$ by a square function. This simplified  description allows  further analytical progress while preserving the same qualitative behavior. Since the Debye regime can be neglected,  we choose:
\[
D(\omega)=\left\{
\begin{array}{l l}
\frac{1}{\omega_c-\omega^*} & \quad \omega^*\leq\omega\leq\omega_c\quad \delta z\leq0\\
\frac{1}{\omega_c} & \quad 0\leq\omega\leq\omega_c\quad \delta z>0.\\
\end{array} \right.
\]
Considering $\omega^*=\frac{\vert\delta z\vert}{z_c}\omega_c$, the cut-off frequency $\omega_c\sim1$ is  the only fitting parameter of the simplified continuum model. Rescaling as $\alpha\rightarrow\alpha \omega^2_c$, we obtain that all the thermodynamic functions depend uniquely on $\alpha=\frac{z_{\rm w}k_{\rm w}}{d k \omega_c^2}$,  $T$ and $\delta z$. In particular, the specific heat is:
 
\begin{equation}
c(T,\delta z,\alpha)=\left\{
\begin{array}{l l}
\frac{z_c}{4z}\frac{\sqrt{\alpha(1+1/T)}}{(1+T)^2}\left[\arctan(\frac{1}{\sqrt{\alpha(1+1/T)}})\right. &\\
\quad+\frac{\sqrt{\alpha(1+1/T)}}{1+\alpha(1+1/T)}+\arctan(\frac{\delta z/z_c}{\sqrt{\alpha(1+1/T)}})\\
\quad+\left.\frac{\delta z}{z_c}\frac{\sqrt{\alpha(1+1/T)}}{\delta z^2/z_c^2+\alpha(1+1/T)}\right] & \delta z\leq0\\
\\
\frac{z_c}{4z}\frac{\sqrt{\alpha(1+1/T)}}{(1+T)^2}\left[\arctan(\frac{1}{\sqrt{\alpha(1+1/T)}})\right. &\\
\quad+\left.\frac{\sqrt{\alpha(1+1/T)}}{1+\alpha(1+1/T)}\right]+\frac{\delta z}{2z}\frac{1}{(1+T)^2} & \delta z>0.\\
\end{array} \right.
\nonumber
\end{equation}

We compute the jump of specific heat at the Kautzman temperature, where the entropy vanishes $s(T_K,\delta z,\alpha)=0$.  In the continuous limit, the equations for $T_K$ can be approximated by:

\[
\ln2\approx \frac{z_c}{2z}\left[\ln(1+\frac{\alpha}{T_K})+2\sqrt{\frac{\alpha}{T_K}}\arctan\sqrt{\frac{T_K}{\alpha}}\right] -\frac{\delta z}{2z}\theta(\delta z)\ln{T_K},
\]
where  the conditions  $\vert\delta z\vert\ll z_c$ and $\alpha\sim k_{\rm w}/k\ll1$ have been used. There is no simple analytical expression for $T_K$, however,  one can  observe the existence of two asymptotic regimes: $T_K\sim\alpha$ for $\delta z\ll \vert1/\ln{\alpha}\vert$ and $T_K\sim2^{-2z/\delta z}$ for $\delta z\gg \vert1/\ln{\alpha}\vert$. Then the specific heat at the transition temperature is given by:

\[
c(T_K,\delta z,\alpha)\sim\left\{
\begin{array}{l l}
\frac{z_c}{4z}\frac{\pi}{2} & \delta z\ll \vert1/\ln{\alpha}\vert\\
\\
\frac{\delta z}{2z} & \delta z\gg \vert1/\ln{\alpha}\vert .\\ 
\end{array} \right.
\]
From these asymptotic  behaviors one gets that the  specific heat display a non-monotonous behavior with coordination, with a minimum whose position scales as $\delta z\sim \vert1/\ln{\alpha}\vert$.

\end{article}

\end{document}